# Could AI change the scientific publishing market once and for all?

**Wadim Strielkowski**

University of California, Berkeley, Giannini Hall, CA 94720 Berkeley, United States, e-mail: strielkowski@berkeley.edu

Prague Business School, Werichova 29, 15200 Prague, Czech Republic, e-mail: strielkowski@pbs-education.cz

**Abstract:** Artificial-intelligence tools in research like ChatGPT are playing an increasingly transformative role in revolutionizing scientific publishing and re-shaping its economic background. They can help academics to tackle such issues as limited space in academic journals, accessibility of knowledge, delayed dissemination, or the exponential growth of academic output. Moreover, AI tools *could* potentially change scientific communication and academic publishing market as we know them. They can help to promote Open Access (OA) in the form of preprints, dethrone the entrenched journals and publishers, as well as introduce novel approaches to the assessment of research output. It is also imperative that they *should* do just that, once and for all.

**Keywords:** AI-assisted writing, academic publishing, ChatGPT, scientific publishing market, open access

## 1. Introduction

Artificial intelligence (AI) tools in academic publishing did not quite take the world by surprise (Van Noorden and Perkel, 2023; Lund et al., 2023). Even though they existed in some form before, it was ChatGPT, an AI language model developed by OpenAI that enables interactive conversations with users on various topics, released and made publicly available in November 2022 that might now change the rules of the game.

From an economic perspective, it is evident the system of academic publishing is highly unsustainable and financially imprudent (Molchanova et al., 2017; Strielkowski and Gryshova, 2018; Strielkowski and Chigisheva, 2018; or Boufarss and Harviainen, 2021). Moreover, as Strielkowski (2017) puts it, many academics struggle with the thought that the results of scientific research can be bought and sold, while the whole mechanism of academic publishing operates on a basis when the buyer pays twice for what the seller offers. The academic publishing system, established in the aftermath of World War II and largely unchanged to this day, follows a sequence of steps:

- Governments allocate funds to support scientific research;
- Scientists conduct their research and disseminate their findings through academic journals;
- Governments provide additional funding to publish these scientific results in academic journals, along with covering the subsequent access through numerous databases.

Due to limited space in journals and strict criteria for publication, many valuable research studies often go unnoticed or unpublished. The emergence of AI tools like ChatGPT is changing academic publishing by providing an alternative platform for researchers to share their work more efficiently and inclusively (Conroy, 2023). ChatGPT can assist scientists in writing manuscripts by suggesting relevant literature references or helping them refine their arguments through real-time feedback. By leveraging AI technologies like ChatGPT, researchers can streamline their writing processes and collaborate more effectively.

**2. Benefits that the AI-assisted writing for scientific publishing**

AI tools offer unprecedented opportunities to accelerate the dissemination of research, enhance collaboration among scientists, and improve the overall quality and accessibility of published works. One key aspect where AI tools are revolutionizing scientific publishing is in automating various labour-intensive tasks involved in manuscript preparation and review processes (Parrilla, 2023). For instance, AI-powered language models like ChatGPT can assist researchers in writing high-quality manuscripts by suggesting improvements to grammar, style, clarity, and conciseness. Peer review traditionally relies on human reviewers who assess manuscripts for originality, methodology, significance of findings, and adherence to ethical guidelines. However, this process is often time-consuming and subject to biases or inconsistencies. AI tools can aid in automating parts of this process by identifying potential conflicts of interest among reviewers or even providing preliminary assessments based on existing literature. By doing so, they help expedite manuscript evaluation while maintaining objectivity and transparency. Figure 1 shows the benefits of the AI-assisted writing for scientific publishing.

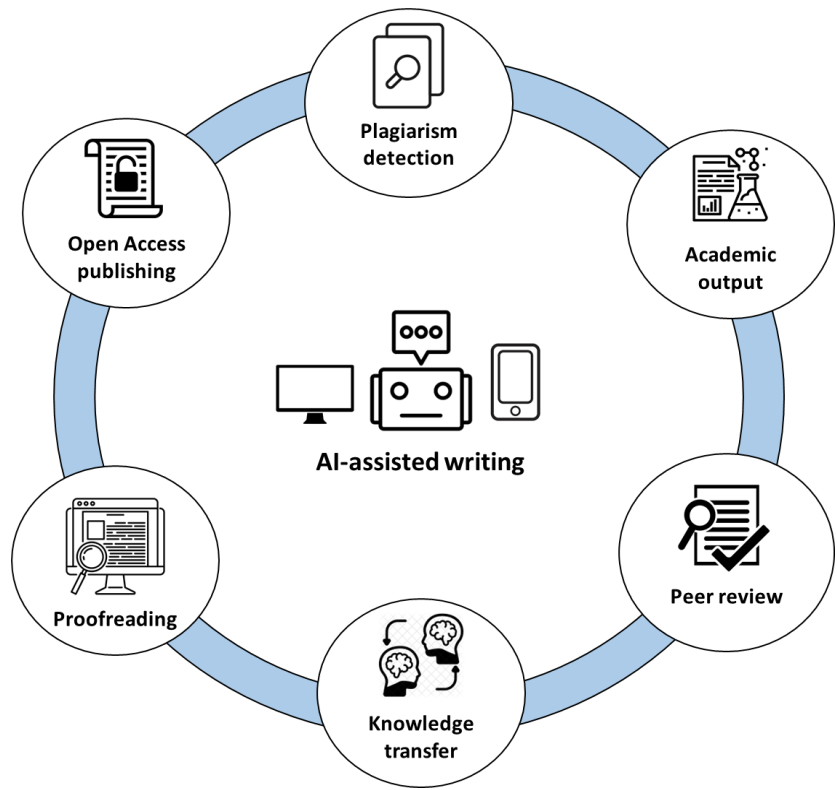

**Figure 1.** Benefits of the AI-assisted writing for scientific publishing
Source: Own results

In addition to improving efficiency within traditional publishing workflows, AI tools can also enable new modes of collaboration among scientists across different disciplines or geographical locations. For example, online platforms using AI algorithms can match researchers with similar interests or expertise for potential collaborations on specific projects or publications. These platforms facilitate networking opportunities that were previously limited by physical boundaries or lack of awareness about each other's work. Another significant contribution of AI tools in revolutionizing scientific publishing is their ability to enhance the accessibility and discoverability of research outputs. AI-powered search algorithms can help researchers find relevant literature more efficiently, saving time and effort. Additionally, AI tools can generate summaries or visual representations of scientific papers, making them more accessible to a broader audience that may not have the necessary domain expertise to understand complex technical jargon.

The effectiveness of AI in academic publishing is contingent upon the quality of the training data. Presently, only two comprehensive databases exist for this purpose: Scopus (owned by Elsevier) and Web of Science (owned by Clarivate). This exclusivity positions these academic publishing giants uniquely to develop dependable editorial AI tools. Journals not using the manuscript submission systems such as Editorial Manager or ScholarOne might not access such technologies unless Elsevier or Clarivate offer API licensing to other platforms. Here, Elsevier might be more inclined to expand its AI services, given its lesser reliance on platform-specific revenue compared to Clarivate's ScholarOne. Time will show how Scopus and Web of Science are going to adapt to the AI-powered academic publishing but the question in now whether they would but rather how quickly they do that.

## 3. Human contribution in AI-powered research

While AI tools offer numerous benefits, their implementation must be guided by a set of ethical principles to ensure responsible and transparent scientific publishing practices. One primary concern is the potential bias embedded within AI algorithms. The training data used to develop these models may inadvertently perpetuate biases present in the scientific literature itself. Transparency is another critical ethical consideration when using AI tools for scientific publishing. As ChatGPT generates text based on patterns learned from vast amounts of data, it becomes challenging to ascertain how decisions are made within its algorithms. Users should be informed about any limitations or potential biases associated with these tools so that they can interpret their outputs accordingly. It is necessary to strike a balance between providing users with understandable explanations without compromising proprietary information.

Furthermore, a question arises here: "Is human contribution irreplaceable when the AI-powered research is involved?". One can argue that this should not necessarily be the case. Table 1 features some examples of the human contribution in academic research being irreplaceable by any AI.

**Table 1.** Examples of irreplaceable human contributions in AI-powered research

| Field of research | Human contribution |
|---|---|
| Education research | Trust, empathy, and a sense of connection for eliciting authentic responses from students |
| Medical research (babies growth study) | Ability to take measurements and soothe babies during research |
| Creativity research | Genuine stupidity as a human characteristic to enable creativity |
| Drug testing and development | Capability of making mistakes and stumbling upon serendipitous discoveries |
| Agricultural research (grain production) | Evaluation of aroma, taste, and texture of cooked grains for decision-making |
| Climate change mitigation strategies | Ethical and social evaluation, intuition, and creativity in interdisciplinary solutions |
| Food insecurity research | Intercultural savvy, relationships, and insights into local wildlife and geopolitical events |
| Biodiversity research | Data collection in the field and raw data gathering for species population changes analysis |
| Medical research (pathology) | Accurate data annotation, interpretation, unbiased sourcing, and diversity understanding |

| Climate crisis solutions | Understanding human behaviour and decision-making to build effective climate solutions |
| Smart electricity network design | Human unpredictability and irrationality to adjust energy and power system design and ensure user acceptance |
| Pharmacological research and development | Human contribution for product acceptance and compliance with governmental regulations |

Source: Own results based on Heim et al. (2023)

**4. Implications for academic publishing**

In his last book "Brief Answers to the Big Questions" the late theoretical physicist and cosmologist Stephen Hawking noted that there was no sign of scientific and technological development dramatically slowing down and stopping in the nearest future even though the present rate of growth cannot continue for the next millennium. If the exponential growth is continued, there would be ten or more papers in the field of Theoretical Physics published each second with no one to read them (Hawking, 2018).

One of the most promising prospects for ChatGPT and AI tools in scientific publishing is their ability to enhance the peer review process. Peer review plays a crucial role in ensuring the quality and integrity of scientific research, but it can be time-consuming and subjective. With AI-powered tools, reviewers could receive automated assistance in assessing manuscripts. In addition, in the nearest future, AI might take over the editors' and reviewers' responsibilities and peer review and rate the academic output all by itself thus removing the "academic bottleneck" of the delayed peer reviews caused by the search for potential reviewers, administrative burden, as well as time constraints. It becomes clear that AI-assisted writing could and should change scientific publishing market once and for all. It can help to dethrone the entrenched journals and oligopolistic publishers, remove paywalls and library subscriptions, as well as introduce novel technology-driven approaches to the assessment of research output (such as papers evaluated by AI system on a 100-point scale and then published as open access preprints).

In general, there might be three possible scenarios of how the AI would go along in academic writing and publishing in the years to follow:

1. "Enhanced Academia": It would morph academic writing and publishing into something like the "Enhanced Games", an international sports event planned to be held in December 2024, where the athletes are not going be subject to drug testing (Stazicker, 2023).
2. "Co-pilot"" It would remain a sort of a helping hand, when journals and publishers would allow academics to use AI for drafting their papers or generating ideas but would strictly prohibit to use AI-generated content in the submitted and published papers. There will be AI-detection software tools available, probably embedded into the current established and well-known plagiarism check tools.
3. "AI-free": any use of AI would be banned in academic writing and publishing.

**5. Conclusions**

Overall, the use of AI tools (such as ubiquitous ChatGPT as well as others) has already begun to revolutionize scientific publishing market, but their potential for further transformation remains immense and for the most part yet unexplored. As technology continues to advance and improve, these tools are poised to play an even more significant role in shaping the landscape of scientific publishing in the future.

The integration of AI into academic publishing signifies a pivotal change. Researchers need to adapt to this emerging landscape, favouring concise, clear, and direct writing styles compatible with

AI analysis. This trend underscores the growing influence of AI in shaping the future of academic publishing, with significant implications for editorial processes and the broader industry dynamics. As AI continues to permeate the scientific publishing market, stakeholders and academics alike must remain agile and responsive to these technological advancements.

**References**

- Boufarss, M., & Harviainen, J. T. (2021). Librarians as gate-openers in open access publishing: A case study in the United Arab Emirates. The Journal of Academic Librarianship, 47(5), 102425. https://doi.org/10.1016/j.acalib.2021.102425
- Conroy, G. (2023). How ChatGPT and other AI tools could disrupt scientific publishing. Nature, 622(7982), 234-236. https://doi.org/10.1038/d41586-023-03144-w
- Hawking, S. (2018). Brief Answers to the Big Questions, 1st ed., New York: Bantam Books
- Heim, A., Bharani, T., Konstantinides, N., Powell, J., Srivastava, S., Cao, X., Agarwal, D., Waiho, K., Lin, T., Virgüez, E., Strielkowski, W. and Uzonyi, A. (2023). AI in search of human help. Science, 381(6654), 162-163. https://doi.org/10.1126/science.adi8740
- Lund, B. D., Wang, T., Mannuru, N. R., Nie, B., Shimray, S., & Wang, Z. (2023). ChatGPT and a new academic reality: Artificial Intelligence-written research papers and the ethics of the large language models in scholarly publishing. Journal of the Association for Information Science and Technology, 74(5), 570-581. https://doi.org/10.1002/asi.24750
- Molchanova, A., Chunikhina, N., & Strielkowski, W. (2017). Innovations and academic publishing: who will cast the first stone? Marketing & Management of Innovations, 4, 40-48. https://doi.org/10.21272/mmi.2017.4-03
- Parrilla, J. (2023). ChatGPT use shows that the grant-application system is broken. Nature, 623(7986), 443. https://doi.org/10.1038/d41586-023-03238-5
- Stazicker, M. (2023). A doping free-for-all Enhanced Games calls itself the answer to doping in sports. Available at: https://edition.cnn.com/2023/10/30/sport/enhanced-games-olympics-doping-spt-intl/index.html
- Strielkowski, W. (2017). Will the rise of Sci-Hub pave the road for the subscription-based access to publishing databases? Information Development, 33(5), 540-542. https://doi.org/10.1177/0266666917728674
- Strielkowski, W., & Chigisheva, O. (2018). Research Functionality and Academic Publishing: Gaming with Altmetrics in the Digital Age. Economics & Sociology, 11(4), 306-316. https://doi.org/10.14254/2071-789X.2018/11-4/20
- Strielkowski, W., Gryshova, I. (2018). Academic Publishing and Predatory Journals. Science and Innovation, 14(1), 5-12.
- Van Noorden, R., & Perkel, J. M. (2023). AI and science: what 1,600 researchers think. Nature, 621(7980), 672-675. https://doi.org/10.1038/d41586-023-02980-0